\documentclass[runningheads]{llncs}
\usepackage[T1]{fontenc}
\usepackage{graphicx}
\usepackage{multirow}%
\usepackage{amsmath,amsfonts,amssymb}
\usepackage{mathrsfs}%
\usepackage[title]{appendix}%
\usepackage{xcolor}
\usepackage{textcomp}%
\usepackage{manyfoot}%
\usepackage{booktabs}%
\usepackage{algorithm}%
\usepackage{algorithmicx}%
\usepackage{algpseudocode}%
\usepackage{listings}%
\usepackage{url}
\usepackage[breaklinks=true, colorlinks=true, linkcolor=blue, citecolor=blue, urlcolor=blue]
{hyperref}
\usepackage{orcidlink}

\newcommand\blfootnote[1]
{%
  \begingroup
  \renewcommand\thefootnote{}\footnote{#1}%
  \addtocounter{footnote}{-1}%
  \endgroup
}
\begin{document}
\title{Learning in Blocks: A Multi Agent Debate Assisted Personalized Adaptive Learning Framework for Language Learning}
\titlerunning{\textit{Learning in Blocks}}
\author{Authors}

\makeatletter
\newcommand{\equalfootnote}{\textsuperscript{\@fnsymbol{1}}}
\makeatother

\author{
Nicy Scaria\inst{1}\orcidlink{0009-0004-8699-0312}\thanks{These authors contributed equally.}
\and
Silvester John Joseph Kennedy\inst{1,2,3}\orcidlink{0009-0004-2788-6283}\equalfootnote
\and
Deepak Subramani\inst{1}\orcidlink{0000-0002-5972-8878}
}

\authorrunning{N. Scaria et al.}

\institute{Indian Institute of Science, India \and
Talking Yak, India \and Indian Institute of Technology Patna, India\\
\email{nicyscaria@iisc.ac.in}}
%
% \institute{Affiliation \\
% \email{email}}
\maketitle              % typeset the header of the contribution
\begin{abstract}
Most digital language learning curricula rely on discrete-item quizzes that test recall rather than applied conversational proficiency. When progression is driven by quiz performance, learners can advance despite persistent gaps in using grammar and vocabulary during interaction. Recent work on LLM-based judging suggests a path toward scoring open-ended conversations, but using interaction evidence to drive progression and review requires scoring protocols that are reliable and validated. We introduce \textit{Learning in Blocks}, a framework that grounds progression in demonstrated conversational competence evaluated using CEFR-aligned rubrics. The framework employs heterogeneous multi-agent debate (HeteroMAD) in two stages: a scoring stage where role-specialized agents independently evaluate Grammar, Vocabulary, and Interactive Communication, engage in debate to address conflicting judgments, and a judge synthesizes consensus scores; and a recommendation stage that identifies specific grammar skills and vocabulary topics for targeted review. Progression requires demonstrating 70\% mastery, and spaced review targets identified weaknesses to counter skill decay. We benchmark four scoring and recommendation methods on CEFR A2 conversations annotated by ESL experts. HeteroMAD achieves a superior score agreement with a 0.23 degree of variation and recommendation acceptability of 90.91\%. An 8-week study with 180 CEFR A2 learners demonstrates that combining rubric-aligned scoring and recommendation with spaced review and mastery-based progression produces better learning outcomes than feedback alone.

\keywords{Multi-Agent Debate \and Personalized Adaptive Learning \and Mastery Based Learning.}
\end{abstract}

\blfootnote{This is a preprint. The Version of Record of this contribution will be published in International Conference on Artificial Intelligence in Education (LNAI), and will be available shortly.}

% \blfootnote{This is a preprint. The Version of Record of this contribution is published in International Conference on Artificial Intelligence in Education (LNAI,volume 14830), and is available online at \url{https://doi.org/10.1007/978-3-031-64299-9_12}}
%
%
%
\section{Introduction}

Digital language learning curricula typically rely on discrete-item quizzes that test recall of grammar rules or vocabulary definitions, creating a disconnect between assessment formats (isolated recall tasks) and and what constitutes communicative proficiency (applied performance in context). When progression is driven by quiz performance, learners can advance through content while gaps in applying Grammar, Vocabulary, and Interactive Communication persist, producing a \textit{Swiss cheese} learning \cite{khan2012one} pattern. Large language models have enabled conversational AI applications for language learning \cite{kasneci2023chatgpt}, allowing learners to engage in open-ended dialogue practice. However, using open-ended spoken interaction for assessment and progression requires scoring protocols that are reliable and validated against expert judgment. In the absence of such validation, platforms often default to quiz-based signals that are straightforward to standardize and score at scale, even when those signals do not capture whether learners can apply skills during interaction. Recent work on rubric-based LLM judging \cite{zheng2023judging,li2024generation} suggests a path toward scoring open-ended interaction more directly, creating an opportunity to ground progression decisions in demonstrated conversational competence rather than quiz performance.

This challenge is especially acute for conversational proficiency, which requires coordinated control of grammatical accuracy, lexical appropriateness, and interactional success across multiple turns \cite{hashemi-etal-2024-llm}. In language learning contexts, CEFR provides widely adopted descriptors that operationalize these dimensions through rubric-aligned criteria for Grammar, Vocabulary, and Interactive Communication \cite{cefr_companion}. Applying such rubrics to open-ended conversations requires context-dependent judgments that weigh multiple aspects of performance simultaneously. At the same time, studies of LLM-as-a-Judge protocols report systematic pitfalls, including position bias, verbosity bias, and overconfident single-model judgments \cite{scaria2025evalyaks}, which complicate the use of automated scores for instructional decisions.

Beyond reliable scoring, operationalizing mastery-based progression for conversational practice raises two additional challenges. First, what constitutes sufficient evidence of mastery when success cannot be reduced to item correctness? Learners may demonstrate knowledge of grammatical rules in isolation, yet struggle to use them accurately in conversation \cite{ellis2011implicit}. Second, how can these systems support skill maintenance over time? Even after learners show strong performance, skills can decay without targeted re-elicitation and reinforcement \cite{ebbinghaus1913memory}.

We address these challenges through \textit{Learning in Blocks}, which links progression and review to evidence from conversations evaluated with CEFR-aligned rubrics. Conversational scoring and recommendations are produced using heterogeneous multi-agent debate (HeteroMAD). For each conversation, the system outputs CEFR-aligned scores for Grammar, Vocabulary, and Interactive Communication, along with feedback and ranked recommendations for Grammar skills and Vocabulary topics to revisit. Recommendations are converted into targeted review lesson groups, and progression to the next Concept Block occurs only after the learner meets a 70\% mastery criterion on the block-aligned evaluation.

Our contributions are threefold: (1) we benchmark multiple conversation-scoring mechanisms and select HeteroMAD for CEFR-aligned scoring and recommendations; (2) we formalize a curriculum structure that organizes Grammar, Vocabulary, and Conversation into aligned lesson groups, so that each conversation lesson is designed to elicit the grammar and vocabulary taught in the surrounding instruction; and (3) we run an 8-week learner study with three cohorts to isolate the impact of \textit{Learning in Blocks} framework.

\section{Related Work and Background} 

\subsection{Generative AI for Conversational Assessment and Diagnosis}

LLMs are increasingly used for open-ended language practice and feedback \cite{kasneci2023chatgpt}, but dialogue assessment differs from objective grading because it requires rubric-based judgments of quality, appropriateness, and task success that are context-dependent and span multiple turns. Conversational proficiency is inherently multi-dimensional, involving tradeoffs across grammatical control, lexical appropriateness, and interactional success \cite{hashemi-etal-2024-llm}. In language learning contexts, CEFR provides widely used descriptors that motivate rubric-aligned dimensions such as Grammar, Vocabulary, and Interactive Communication (IC) \cite{cefr_companion}. These challenges have motivated `LLM-as-a-Judge' approaches for scoring open-ended outputs \cite{zheng2023judging,li2024generation}, but prior work also reports systematic pitfalls such as position and verbosity biases, underscoring the need for careful protocol design and explicit validation \cite{scaria2025evalyaks}.

Self-Consistency and Self-Refine are common test-time strategies for improving reliability by aggregating multiple generations or iteratively revising a single output \cite{wang2023selfconsistency,madaan2023self}. Multi-agent debate (MAD) extends this idea by distributing evaluation across multiple agents that propose and critique judgments before a final decision is made \cite{improvingfactuality}. Previous evidence suggests that debate-style methods do not consistently outperform strong baselines in well-specified objective tasks \cite{choi2025debate}, but rubric-based dialog assessment may benefit from multiple perspectives, as it requires judgment across dimensions. Prior work has also explored multi-agent architectures for conversation-based assessment \cite{hou2025llm}. However, validated comparisons of diagnostic mechanisms for CEFR-aligned dialogue scoring and the relevance of generated recommendations remain limited. We benchmark Self-Consistency, Self-Refine, and both homogeneous and heterogeneous MAD variants, testing whether role prompting alone or combining complementary model perspectives yields more stable rubric-aligned scores and more actionable recommendations.

\subsection{Mastery-Based Progression}

Mastery-based learning is built on the premise that learning outcomes must remain constant, allowing variable time for individual learners so that learners progress only after demonstrating a clear performance criterion \cite{bloom1968learning,guskey2010lessons}. However, many digital language curricula operationalize mastery using discrete-item quizzes that primarily test recall of grammar rules or vocabulary definitions. While such assessments efficiently measure declarative knowledge, they do not necessarily predict whether learners can apply those concepts in sentence building or real-time conversation \cite{ellis2011implicit}, leading to weak transfer where surface progression masks persistent application gaps \cite{koedinger2012knowledge}. A central challenge for mastery-based progression in open-ended language practice is, therefore, defining mastery evidence when success is not reducible to item correctness. Prior adaptive systems have explored competency-based progression using knowledge tracing and Item Response Theory-style ability estimates \cite{gervet2023kt_survey,shen2024survey,yuhana2024enhancing}, but these approaches predominantly model item or response-level performance rather than applied performance in open-ended interaction. Recent work has examined mastery-based progression and thresholds in online learning environments \cite{matayoshi2025using,pardos2023oatutor}, yet robust, rubric-aligned criteria for conversational mastery remain comparatively under-specified in language learning settings \cite{liu2025evaluating,karatay2025exploring}. These limitations motivate mastery criteria grounded in application of skills in conversations, coupled with targeted remediation.

\subsection{Retention and Skill Maintenance}

Spaced repetition is widely used to support long-term retention of discrete language items, and its benefits are commonly explained through the spacing effect and the forgetting curve \cite{cepeda2006distributed,ebbinghaus1913memory}. However, item retention does not necessarily imply the ability to apply knowledge in context. Learners may recall vocabulary or rules in isolation, yet fail to deploy them accurately during sentence production or conversation. Consequently, we use conversation-derived recommendations to trigger brief, targeted review activities. These reviews focus on restoring performance and can be verified through subsequent mastery checks.

Prior work has examined conversation practice and assessment, mastery-based progression, and spaced repetition. However, these components are often studied in isolation, and validated links between open-ended interaction evidence, rubric-aligned scoring, and targeted remediation remain limited. Building on this gap, we introduce \textit{Learning in Blocks}, which uses CEFR-aligned scoring and recommendations from learner conversations to drive targeted review lesson groups, requiring learners to satisfy a mastery criterion before advancing to the next Concept Block.

% \section{Terminology and Formulation} \label{terminology}

\section{Curriculum Structure}

Let $(CB_t)$ denote the curriculum block at instructional index (t). Each block consists of instructional activities organized around three components: Grammar (G), Vocabulary (V), and Conversation (C). All lessons share a common instructional structure, including video-based instruction, flashcards, controlled practice, and end-of-lesson assessments of objective and subjective types. Grammar lessons emphasize rule acquisition and sentence construction; Vocabulary lessons introduce level-appropriate lexical items selected to support communicative tasks; and Conversation lessons operationalize grammatical rules and lexical knowledge through contextualized use. The pedagogical design and learning objectives are derived from the CEFR Companion Volume \cite{cefr_companion}.

Each curriculum block $(CB_t)$ is modeled as a sequence of $n$ GVC groups,$$CB_t = \bigl(GVC_{t,1}, GVC_{t,2}, \dots, GVC_{t,n}\bigr),$$ 
where each group $GVC_{t,i} = (G_{t,i}, V_{t,i}, C_{t,i})$ enforces a fixed instructional progression from grammatical rule presentation to lexical preparation and finally to communicative application. The ordering constraint, $G_{t,i} \rightarrow V_{t,i} \rightarrow C_{t,i}$, guarantees that structural and lexical knowledge precede use. Let $\mathcal{R}(G_{t,i})$ denote the set of grammatical rules and sentence-level constructions explicitly taught in the grammar lesson, $G_{t,i}$, and let $\mathcal{W}(V_{t,i})$ denote the set of lexical items explicitly introduced in the vocabulary lesson, $V_{t,i}$. For each conversation lesson, $C_{t,i}$, a guided-practice task, $T_{t,i}$, is specified, and its minimal lexical requirements are given by $\mathrm{LexReq}(T_{t,i})$. The curriculum enforces the coverage condition,

$$\mathrm{LexReq}(T_{t,i}) \subseteq \mathcal{W}(V_{t,i}) \cup \mathcal{W}_{\mathrm{prior}}(t,i),$$ where $\mathcal{W}{_\mathrm{prior}}(t,i)$ denotes vocabulary introduced earlier in the curriculum. Finally, conversation lessons operationalize the combined grammatical rules and lexical sets into contextualized communicative use, ensuring that task completion depends only on previously introduced instructional content.

For instance, at the A2 level (Spoken Interaction), one selected Interaction activity is `Goal-oriented co-operation', defined as the ability to communicate in simple and routine tasks using simple phrases to ask for and provide things. In this instantiation, the Vocabulary lessons introduce level-appropriate items for everyday coordination (e.g., time expressions, sequencing words, and task-related nouns and verbs), while the Conversation lessons use guided scenarios such as organising a small event, planning a simple schedule, or completing a shared task, consistent with CEFR A2 descriptors \cite{cefr_companion}.

\section{The Learning in Blocks Framework}

\begin{figure}[ht]%
\centering
\includegraphics[width=0.8\textwidth]{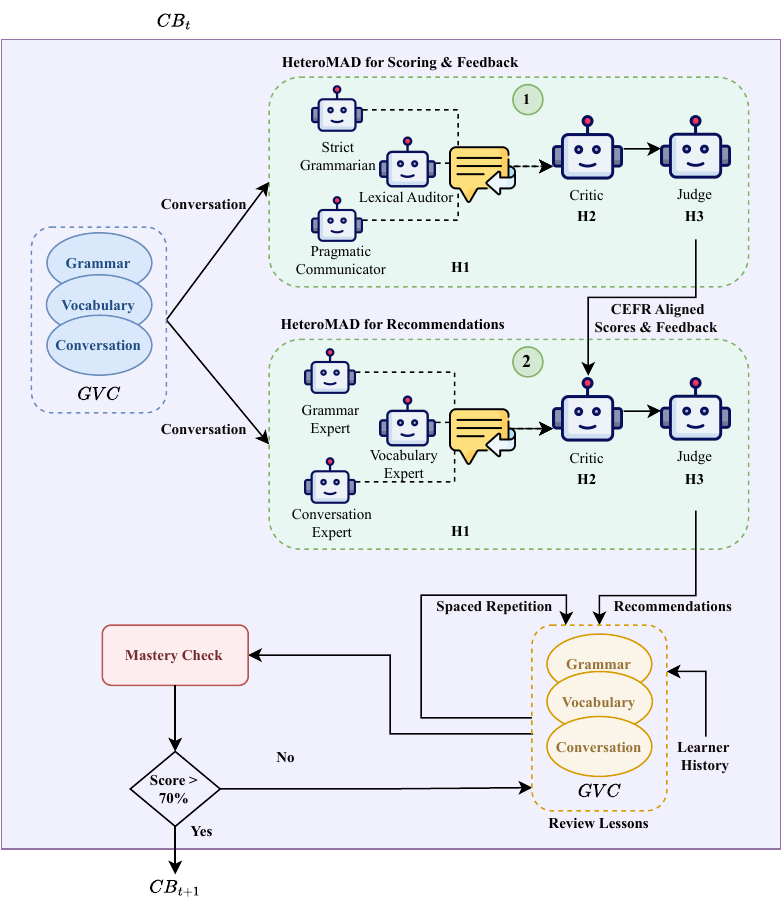}
\caption{\textit{Learning in Blocks} framework within concept block $CB_t$. HeteroMAD produces CEFR-aligned scores and recommendations, triggering review lessons delivered via spaced repetition. Progression to $CB_{t+1}$ requires mastery attainment}\label{block_diagram}
\end{figure}

Our framework  shown in Figure~\ref{block_diagram} orchestrates instruction, evaluation, feedback, and review/remediation using spaced repetition, and it recommends new lesson groups based on demonstrated mastery. Instruction is delivered through structured Concept Blocks that introduce and practice targeted skills. Evaluation happens at the end of every lesson to assess performance across grammar, vocabulary, and communication. Feedback is shared immediately after evaluation, highlighting strengths and clearly stating areas for improvement. Review and remediation are then assigned to counter the forgetting curve, using spaced repetition to reinforce weak or unstable skills over time. This component also connects forward into all future Concept Blocks, continuously updating what the learner practices and when based on evolving mastery. After learners complete all curated lesson groups, a comprehensive mastery check is conducted; using Heterogeneous MAD, we then recommend review lesson groups until the user achieves 70\% mastery.

\subsection{MAD-Based Scoring and Recommendation}

The HeteroMAD Pipeline shown in Figure~\ref{block_diagram} operates in two sequential stages: a scoring stage that produces CEFR-aligned scores, followed by a recommendation stage that generates prioritized recommendations. Both stages use the same three-phase architecture (H1-H2-H3).

The scoring pipeline proceeds in three phases. In H1, three specialized agents independently analyze the conversation using the Cambridge CEFR A2 Key Spoken Interaction rubric \cite{english2024a2}: Agent A (Strict Grammarian) focuses on grammatical accuracy, Agent B (Lexical Auditor) evaluates vocabulary precision, and Agent C (Pragmatic Communicator) assesses IC. Each agent produces scores (0-5) for all three dimensions, along with a rationale that emphasizes their perspective. In H2, a critic facilitates a debate where each agent reviews peer assessments and refines its scores and rationale by addressing conflicting judgments. In H3, a judge synthesizes both initial and refined assessments to produce consensus scores for Grammar, Vocabulary, and IC, along with actionable feedback.

% The recommendation pipeline operates on the scoring stage output. In H1, three specialized agents independently propose review recommendations: Grammar Expert, Vocabulary Expert, and Conversation Expert. Each agent receives the conversation and the complete curriculum of available grammar skills and vocabulary topics, and produces recommendations for up to 2 grammar skills and up to 2 vocabulary topics, along with observations of what went well and what needs development (if any). The agent also provides descriptive, actionable feedback to improve IC (for example, clarity, turn-taking, responsiveness, and appropriate follow-up questions). In H2, agents review peer recommendations and reflect on whether to retain or revise their selections. In H3, a judge identifies consensus choices and resolves disagreements, producing a ranked list of up to the top-2 Grammar Skills and the top-2 Vocabulary Topics. The benchmarking process that led to the selection of HeteroMAD for scoring and recommendation is described in Section~\ref{subsec:diagnostic_benchmark}.

In the recommendation pipeline, H1 includes three specialized agents: Grammar Expert, Vocabulary Expert, and Conversation Expert, each operating independently to propose review recommendations. Each agent receives the conversation and the complete curriculum of available grammar skills and vocabulary topics, and produces recommendations for up to 2 grammar skills and up to 2 vocabulary topics, along with observations of what went well and what needs development (if any). The agent also provides descriptive, actionable feedback to improve IC (for example, clarity, turn-taking, responsiveness, and appropriate follow-up questions). In H2, a critic facilitates a review of peer recommendations, and each agent decides whether to retain or revise its selections; this is the only stage that uses the scoring stage output. In H3, a judge identifies consensus choices and resolves disagreements, producing a ranked list of up to the top-2 Grammar Skills and the top-2 Vocabulary Topics. The benchmarking process that led to the selection of HeteroMAD for scoring and recommendation is described in Section~\ref{subsec:diagnostic_benchmark}.

In both pipelines, agents are provided with the full set of skills and vocabulary to which the learner has been exposed up to that point during evaluation. Accordingly, for any given $GVC_{t,i}$, we expect the learner to draw not only on the skills and vocabulary introduced in the current learning group, but also on any content introduced previously. To avoid penalizing learners for demonstrating knowledge beyond the current group, both pipelines condition their scoring on cumulative exposure. Moreover, we do not penalize learners for using advanced skills or vocabulary that have not yet been taught; instead, we record such instances to conduct diagnostic bridge assessments (DBAs) between two consecutive learning blocks, $CB_{t}$ and $CB_{t+1}$. Based on the learner’s performance in the DBA, we then personalize the subsequent block, $CB_{t+1}$ for example by removing the corresponding learning group $GVC_{t+1,i}$ . In addition, the model produces a textual scoring rationale, which is then used as a context during the agents’ debate when recommending subsequent review groups.

\subsection{Review and Remediation}

A learner is considered to have mastered the current learning block $CB_t=\bigl(GVC_{t,1},GVC_{t,2},\dots,GVC_{t,n}\bigr)$ if they achieve at least 70\% in every GVC contained in $CB_t$, at which point progression to $CB_{t+1}$ is unlocked. For any completed block in which performance is below 100\%, the system logs the specific unmastered skills and vocabulary responsible for the missed percentage and instantiates a corresponding per-block review lesson group, denoted $GVC_{\text{review},b,r}$, where $b$ indexes the originating block and $r$ indexes the review group derived from that block. If a learner has $k$ prior blocks eligible for review, the set $\{GVC_{\text{review},b,r}\}_{(b,r)\in\mathcal{R}_t}$ is carried forward and appended to subsequent blocks (i.e., $CB_t$ is augmented with these review groups). To counter the Ebbinghaus Forgetting Curve \cite{ebbinghaus1913memory}, the carried-forward review groups are re-tested at spaced intervals, specifically on day 2 and day 5 after a block is passed. Each new block begins by assessing $\{GVC_{\text{review},b,r}\}_{(b,r)\in\mathcal{R}_t}$; the resulting scores contribute to the learner's mastery percentage for the current block and are then used to update $\mathcal{R}_{t+1}$, ensuring that earlier weaknesses are continuously revisited until full mastery is demonstrated.

\section{Methodology and Experimental Setup}\label{sec5:experiments}

\subsection{Benchmarking Evaluation and Recommendation}
\label{subsec:diagnostic_benchmark}

We benchmarked four methods to select a reliable configuration for conversation-based scoring and recommendation. We collected a random set of $N=150$ conversations from learners working towards CEFR A2 and transcribed speech using Whisper V3 Turbo \cite{radford2023robust}. Two English as a Second Language (ESL) experts independently scored each conversation for Grammar, Vocabulary, and IC using CEFR-aligned rubrics on an ordinal 0-5 scale. Inter-rater reliability was assessed using percentage agreement and quadratic weighted Cohen's $\kappa$ \cite{cohen1968weighted}. We report percentage agreement and quadratic weighted Cohen's $\kappa$ \cite{cohen1968weighted}. Disagreements were resolved through discussion to obtain consensus scores, used as ground truth. Experts also agreed on the top-2 Grammar and top-2 Vocabulary targets per conversation, prioritizing prerequisite targets before advanced ones.

\subsubsection{Methods and Model Configurations}

We compared Self-Consistency, Self-Refine, homogeneous MAD (HomoMAD), and heterogeneous MAD (HeteroMAD). In this benchmarking stage, each method operated via a two-stage pipeline: (i) a scoring stage to produce per-dimension scores for Grammar, Vocabulary, and IC from the scoring stage with feedback and (ii) recommendation stage to produce ranked recommendations (top-2 Grammar Skills and top-2 Vocabulary Topics, if applicable) with brief rationales. The exact JSON schema and examples are provided in our repository\footnotemark[1].
\footnotetext[1]{\href{https://github.com/Talking-Yak/learningInBlocks}{GitHub Repository}}

To assess the viability of open-source models and to avoid dependency on proprietary API versions that may be deprecated over time, we instantiated all methods using open-source language models. For Self-Consistency, we generated three independent analyses using Qwen3 30B A3B \cite{yang2025qwen3} (hereafter Qwen) and aggregated their outputs for each stage with and without thinking. Self-Refine iteratively refined a single Qwen analysis per stage (with and without thinking). HomoMAD deployed role-specialized agents, all instantiated with Qwen, followed by a critic and a judge, also instantiated with Qwen, for both the scoring and recommendation stages. For HeteroMAD, we instantiated the three specialists with Qwen, Gemma3 27B Instruct \cite{team2025gemma}, and GPT OSS 20B \cite{agarwal2025gpt}, followed by a critic and judge instantiated with Qwen, for both stages. We selected Qwen as the judge model for all MAD variants to keep the synthesis stage fixed across configurations and to ensure consistent structured output. Using an open-source judge also improves reproducibility and avoids dependency on proprietary API versions that may be deprecated, which can hinder exact replication. The prompts, hyperparameters, and inference settings for all methods are provided in the repository.

We evaluate (i) score agreement with expert consensus and (ii) recommendation quality, and report expert inter-rater reliability, defined as follows.

\begin{enumerate}
\item \textit{Expert Inter-Rater Reliability:} percentage agreement and quadratic weighted Cohen's $\kappa$ for each dimension.
    \item \textit{Score Agreement:} degree of variation \cite{scaria2025evalyaks}, which measures the mean absolute deviation between predicted and expert scores, reported per dimension and overall.
    \item \textit{Recommendation Acceptability:} percentage of predicted recommendations matching the expert top-2 Grammar and top-2 Vocabulary expert recommendation.
\end{enumerate}

\subsection{Learner Assignment and Intervention Design}

The study involved $N=180$ English learners currently working toward CEFR A2 proficiency, all following the same 8-week sequence of Concept Blocks and completing the same conversation lessons. During the first two weeks, all learners received identical instruction and feedback generated via Self-Consistency to establish baseline performance. At the end of Week 2, learners were assessed and randomly assigned to one of three cohorts ($N=60$ each), after which interventions were administered.

The first cohort continued to receive CEFR-aligned scores and feedback generated via Self-Consistency for the remainder of the study. No targeted review lessons were assigned, and progression was not gated by mastery. The second cohort received CEFR-aligned scores and feedback generated via HeteroMAD for the remainder of the study, also without targeted review. The third cohort followed the \textit{Learning in Blocks} framework.

The learning outcomes were evaluated using CEFR-aligned scores for Grammar, Vocabulary and IC calculated for the Concept Block completed in Week~8. Each learner completed multiple conversations in Week~8; we report a cumulative Week~8 score for each dimension by aggregating scores within the block. Scores were recorded on a 0-5 ordinal scale and rescaled to a 0-100 scale for comparability across cohorts.

\section{Results}

\subsection{Benchmarking Results}

Table~\ref{tab:expert_irr} shows inter-rater reliability between the two expert annotators across all three dimensions. The agreement was strong across all dimensions, with percentage agreement ranging from 86.36\% to 90.91\% and quadratic weighted Cohen's $\kappa$ ranging from 0.76 to 0.87, indicating substantial to near-perfect agreement and establishing a reliable ground truth.

\begin{table}[ht]
\centering
\caption{Expert inter-rater reliability for CEFR-aligned scoring}
\label{tab:expert_irr}
\begin{tabular}{c@{\hspace{1em}}c@{\hspace{1em}}c}
\toprule
Dimension & Percentage Agreement & Quadratic Weighted $\kappa$ \\
\midrule
Grammar & 90.91\% & 0.87 \\
Vocabulary & 88.64\% & 0.85 \\
IC & 86.36\% & 0.76 \\
\bottomrule
\end{tabular}
\end{table}

Table~\ref{tab:dov_results} reveals a critical finding: explicit thinking mechanisms substantially degrade rubric-aligned assessment performance. Self-Refine's degree of variation more than doubled from 0.41 to 1.06 when thinking was enabled, with similar degradation in Self-Consistency (0.45 to 0.89). This degradation was most severe for Vocabulary, where Self-Refine with thinking reached 1.27, nearly triple the non-thinking variant, suggesting that rubric-based scoring benefits from direct pattern matching rather than verbose reasoning chains.

\begin{table}[ht]
\centering
\caption{Degree of variation between ground truth and method-generated scores}
\label{tab:dov_results}
\setlength{\tabcolsep}{1.5pt}
\begin{tabular}{ccccccc}
\toprule
Dimension & S-Refine & S-Refine & S-Consistency & S-Consistency & Homo & Hetero \\
& (NT) & (WT) & (NT) & (WT) & MAD & MAD \\
\midrule
Grammar & 0.32 & 0.98 & 0.41 & 0.82 & 0.27 & \textbf{0.11} \\
Vocabulary & 0.43 & 1.27 & 0.48 & 1.00 & 0.64 & \textbf{0.27} \\
IC & 0.48 & 0.93 & 0.48 & 0.86 & 0.39 & \textbf{0.32} \\
\midrule
Average DOV & 0.41 & 1.06 & 0.45 & 0.89 & 0.43 & \textbf{0.23} \\
\midrule
\textit{Closest Match (\%)} & \textit{34.09} & \textit{11.36} & \textit{25.00} & \textit{2.27} & \textit{38.64} & \textit{\textbf{70.45}} \\
\bottomrule
\end{tabular}
\end{table}

Among non-MAD approaches, ensemble methods showed minimal advantage: Self-Refine and Self-Consistency without thinking performed nearly identically (0.41 vs. 0.45). Role specialization in HomoMAD yielded only a marginal improvement (0.43), whereas model heterogeneity in HeteroMAD achieved a degree of variation of 0.23. The `Closest Match' metric reinforces that HeteroMAD produced the most accurate scores in 70.45\% of conversations, nearly double HomoMAD's 38.64\%, demonstrating consistent superiority.

Recommendation quality showed a different pattern within the thinking mechanisms:  thinking improved acceptability. Self-Consistency with thinking achieved 65.91\% acceptability compared to 45.45\% without thinking, suggesting that recommendation generation benefits from explicit reasoning about priorities. Self-Refine with thinking achieved 59.09\% and without thinking achieved 54.55\%. However, even the best non-MAD method remained below HeteroMAD's 90.91\% acceptability. HomoMAD achieved 59.09\%, matching Self-Refine with thinking and indicating that role specialization provides equivalent benefits to iterative refinement for recommendations.

Based on HeteroMAD's combined superiority in score agreement (0.23 degree of variation) and recommendation acceptability (90.91\%), it was selected for deployment in the learner study.

\subsection{Learner Study Results}

\begin{figure}[H]
\centering
\includegraphics[width=0.85\textwidth]{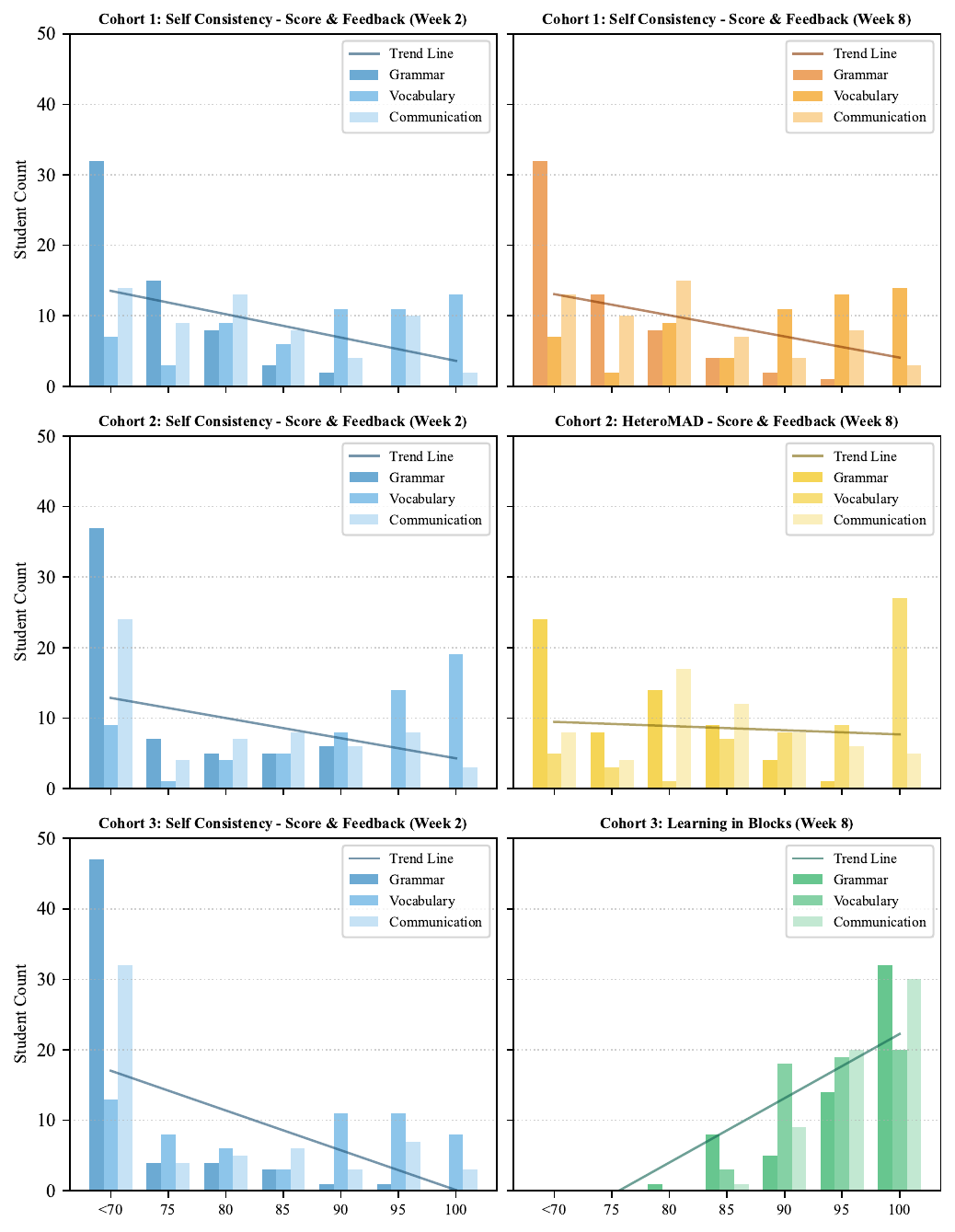}
\caption{Distribution of CEFR-aligned scores at Week 2 (baseline) and Week 8 (post-intervention) for three cohorts ($N=60$ each).}
\label{fig:learner_distributions}
\end{figure}

Figure~\ref{fig:learner_distributions} shows the distribution of CEFR-aligned scores at baseline (Week 2) and post-intervention (Week 8) for Grammar, Vocabulary, and IC across the three cohorts. At baseline, all cohorts received Self-Consistency feedback and showed comparable distributions, with most learners scoring below 70\% across all dimensions.

At Week 8, the interventions produced different learning trajectories. Cohort 3 (\textit{Learning in Blocks}) showed dramatic improvement, with distributions shifting rightward and concentrating in the 85-95\% range, particularly for Grammar. Most learners achieved scores above the 70\% mastery threshold across all dimensions. Cohort 2 (HeteroMAD feedback) demonstrated moderate gains, with distributions shifting toward higher scores but exhibiting greater dispersion than Cohort 3. Cohort 1 (Self-Consistency feedback) showed minimal improvement, with many learners remaining below 70\% and distributions similar to baseline patterns. This progression reveals that superior feedback alone (Cohort 2) produces moderate gains, but pairing it with mastery-based progression and targeted review (Cohort 3) produces substantially stronger outcomes.

Paired t-tests confirmed statistically significant pre-post gains for Cohort 3 across all dimensions (all $p < .001$), with large to very large effect sizes (Cohen's $d = 0.98$-$1.97$). Cohort 2 also showed significant gains across all dimensions, with medium-to-large effects ($d = 0.56$-$1.12$, all $p < .001$), whereas C1 showed smaller gains in Grammar and IC (both $p < .001$), with no significant improvement in Vocabulary ($p = .103$). Between-cohort ANOVAs revealed significant differences in gain scores for Grammar, Vocabulary, and IC (all $p < .001$). Comparisons showed that Cohort 3 significantly outperformed both 2 and 1 across all three dimensions (all $p < .001$, $d = 0.67$-$1.99$), while differences between Cohort 2 and 1 were not significant (all $p > .05$), suggesting that the largest improvements arise when HeteroMAD feedback is combined with structured review and mastery-based progression.

All Week 8 scores are based on Concept Block 7 conversations to ensure a fair comparison across cohorts. Although Cohorts 1 and 2 had completed Block 8 by Week 8, most Cohort 3 learners were at Block 7 (with some reaching Block 8), because progression in this condition required demonstrating mastery before unlocking the next block. As a result, Cohort 3 learners spent more time on earlier content. This additional time is a consequence of the framework, rather than a separate source of variation, since requiring demonstrated mastery before progression is a core part of the intervention design.

\section{Conclusion}

We introduced \textit{Learning in Blocks}, a framework that integrates HeteroMAD-based conversational assessment with mastery-based progression and spaced repetition for CEFR-aligned language learning. Our benchmarking study demonstrated that HeteroMAD outperforms Self-Consistency, Self-Refine, and HomoMAD in both score agreement and recommendation acceptability. The HeteroMAD architecture combines complementary perspectives through role-specialized debate, increasing reliability in rubric-aligned assessment.

In an 8-week study with 180 learners working toward CEFR A2, combining CEFR-aligned assessment with targeted review and mastery-based progression produced better outcomes than feedback alone. Cohort 3 (\textit{Learning in Blocks}) achieved the largest shift toward higher performance across Grammar, Vocabulary, and IC, with no learners falling below the mastery threshold. In comparison, Cohort 2 (HeteroMAD feedback only) showed moderate gains and Cohort 1 (Self-Consistency baseline) exhibited minimal improvement. Although mastery-based progression slowed advancement in Cohort 3, requiring mastery before progression enabled learners to develop stable proficiency and stronger cross-block application rather than advancing with persistent gaps.

Our work addresses a critical gap in adaptive language learning by operationalizing mastery criteria for open-ended conversational practice, where success cannot be reduced to item correctness. By instantiating the multi-agent architecture with open-source models, the framework supports reproducible deployment without reliance on proprietary API versions. Future work should explore longitudinal retention beyond the 8-week intervention period, extend the framework to additional CEFR levels and languages, investigate optimal review scheduling to counter individual forgetting curves, and incorporate audio-modal signals beyond transcripts to enable more comprehensive assessment of pronunciation intelligibility and phonological features in spoken interaction.


\begin{credits}

\subsubsection{\ackname}

We sincerely thank language expert, Cynthia Freeda B, for her support in reviewing and scoring the conversational data used in this work. We also thank the additional anonymous expert reviewers for their careful assessments. Finally, we thank Thomas Latinovich, CEO of Talking Yak, for supporting this research.

\subsubsection{\discintname}
The authors have no competing interests to declare that are relevant to the content of this article.
\end{credits}
\bibliographystyle{splncs04}
\bibliography{references}




% LNCS Homepage, \url{http://www.springer.com/lncs}, last accessed 2023/10/25
% \end{thebibliography}
\end{document}